# Protonation enhancement by dichloromethane doping in low-pressure photoionization

Jinian Shu,[1-3,*] Yao Zou,[1,2] Ce Xu,[1,2] Zhen Li,[1,2] Wanqi Sun,[1,2] Bo Yang,[1,2] Haixu Zhang,[1,2] Peng Zhang,[1,2] and Pengkun Ma[1,2]

Doping has been used to enhance the ionization efficiency of analytes in atmospheric pressure photoionization, which is based on charge exchange. Compounds with excellent ionization efficiencies are usually chosen as dopants. In this paper, we report a new phenomenon observed in low-pressure photoionization: Protonation enhancement by dichloromethane ($CH_2Cl_2$) doping. $CH_2Cl_2$ is not a common dopant due to its high ionization energy (11.33 eV). The low-pressure photoionization source was built using a krypton VUV lamp that emits photons with energies of 10.0 and 10.6 eV and was operated at ~500–1000 Pa. Protonation of water, methanol, ethanol, and acetaldehyde was respectively enhanced by 481.7 ± 122.4, 197.8 ± 18.8, 87.3 ± 7.8, and 93.5 ± 35.5 times after doping 291 ppmv $CH_2Cl_2$, meanwhile $CH_2Cl_2$ almost does not generate noticeable ions itself. This phenomenon has not been documented in the literature. A new protonation process involving in ion-pair and H-bond formations was proposed to expound the phenomenon. The observed phenomenon opens a new prospect for the improvement of the detection efficiency of VUV photoionization.

([1]State Key Joint Laboratory of Environment Simulation and Pollution Control, Research Center for Eco-Environmental Sciences, Chinese Academy of Sciences, Beijing 100085, China

[2]University of Chinese Academy of Sciences, Beijing, China

[3]Shanghai Masteck Environment Co., Ltd, Shanghai, China

[*] Corresponding author. *E-mail address:* jshu@rcees.ac.cn)

## Introduction

Photoionization (PI), a widely used soft ionization technique, is usually coupled to various mass spectrometers for analyzing the chemical composition of samples.[1-5] Atmospheric pressure photoionization (APPI) is a new and highly attractive ionization technique,[6-7] which was developed ~10 years ago with the aim of improving the performance of liquid chromatography-mass spectrometry (LC-MS) for less polar compounds such as polycyclic aromatic hydrocarbons (PAHs). A krypton lamp, which emits VUV photons with energies of 10.0 and 10.6 eV, is usually chosen as the light source in APPI as it is cheap, compact, and robust. Different from classic vacuum photoionization, APPI shows characteristics more similar to those of chemical ionization (CI). The ionization mechanisms commonly observed in CI are also observed in APPI, such as the proton transfer reaction (PTR) and charge exchange. PTR typically takes place when the analyte in question has a higher proton affinity (PA), whereas charge exchange requires that the analyte possesses low ionization energy (IE).

Low-pressure photoionization (LPPI), defined as photoionization running under hundreds to thousands of Pa, has not been used as widely as APPI and conventional vacuum PI. LPPI has characteristics of both vacuum PI and APPI.[8] Apart from molecular ions, protonated ions were found to be dominant for polar compounds. The proton transfer reactions in LPPI can be expressed as follows:

$A + R^+ \rightarrow AH^+ + [R-H]$  (1) 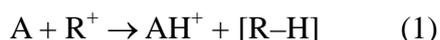

$A^+ + R \rightarrow AH^+ + [R-H]$  (2) 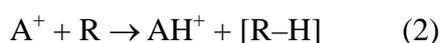

where A represents the analyte molecules and R is the reagent which offers a proton or hydrogen atom. The reagent could be the analyte or solvent molecules.

The use of dopants has been found to be very effective for enhancing the ionization efficiency of analytes[6-7,9-11] in APPI and LPPI[12] via charge exchange:

$D + h\nu \rightarrow D^+ + e$  (3) 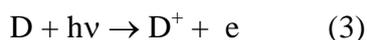

$D^+ + A \rightarrow A^+ + D$  (4) 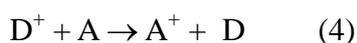

where D and A represent dopant and analyte molecules, respectively. Benzene (IE = 9.24 eV),[13-14] acetone (IE = 9.70 eV),[10,15-17] toluene (IE = 8.83 eV),[6-7,10-12,18-20] and anisole (IE = 8.20 eV)[21] are often employed as dopants due to their excellent photoionization efficiencies under illumination of the krypton lamp. The resulting analyte ions may subsequently react with other molecules via proton transfer. The detection sensitivity could be enhanced by ~100 times via doping.[22] However, these dopants cannot be applied to the detection of methanol ($CH_3OH$, IE = 10.84 eV), ethanol ($C_2H_5OH$, IE = 10.48 eV), and acetaldehyde ($C_2H_4O$, IE = 10.23 eV) due to their higher IEs. Dichloromethane has been chosen as a dopant for characterizing the molecular structures of analytes via secondary ion–molecule reactions, rather than for enhancing ionization efficiency.[23]

Our previous studies revealed that LPPI with a specially designed photoionizer was super sensitive (~1000 counts/ppbv) to many organic compounds.[24-26] However, the LPPI detection efficiency for $CH_3OH$, $C_2H_5OH$, and $C_2H_4O$ is very low due to their low ionization efficiencies. In this paper, we report a new phenomenon: The detection efficiencies of the three small volatile organic compounds (VOCs) can be remarkably enhanced via $CH_2Cl_2$ doping. The results and experimental method are described in the following sections.

**Results**

**Protonation enhancement of water and LPPI mass spectrum of $CH_2Cl_2$**

Water ($H_2O$) is an important protonation agent for PTR mass spectrometry. The IE of water is 12.62 eV, which indicates that it cannot be photoionized directly by the photons emitted from the krypton lamp. However, $H_3O^+$ (m/z 19, 45 counts), $(H_2O)_2H^+$ (m/z 37, 214 counts), and $(H_2O)_3H^+$ (m/z 55, 24 counts) were observed in the LPPI mass spectrum of $N_2$, as shown in Figure 1(A). The concentration of water in the test chamber was <5 ppmv, as a result of impurities in high-purity $N_2$ gas. Protonation of acetonitrile (IE = 12.20 eV) was observed in APPI with a krypton lamp as the VUV light source by Marotta et al. The authors speculate that photon irradiation leads first to the isomerization of acetonitrile molecules, affording species that exhibit IEs <10 eV and that consequently are able to generate photoionization products.[27] The formation mechanism of protonated water and water clusters under illumination of 10.0 and 10.6 eV photons is not clear yet. In view of a tiny amount of $N_2^+$ (m/z 28, 34 counts) observed in Figure 1(A), the photoelectrons in the photoionization region might lead to the formation of protonated water and water clusters. Figure 1(B) shows the mass spectrum obtained after injecting 291 ppmv $CH_2Cl_2$ into the chamber. Surprisingly, the signal intensities of $H_3O^+$, $(H_2O)_2H^+$, and $(H_2O)_3H^+$ increased to $2.92 \times 10^4$, $1.24 \times 10^5$ and $2.29 \times 10^4$ counts, respectively. The signal intensity of protonated water was averagely enhanced by 481.7 ± 122.4 times, measured from three independent measurements. This phenomenon has never been reported.

$CH_2Cl_2$ is a common solvent used in organic analysis. The IE of $CH_2Cl_2$ is 11.33 eV. It cannot be directly ionized by the VUV photons emitted from the krypton lamp. As shown in Figure 1(B), no noticeable ions were produced from direct photoionization of $CH_2Cl_2$. A small mass peaks at m/z 47 is assigned to ethanol residual in the test chamber or minor impurity in the $CH_2Cl_2$ reagent.

**Protonation enhancement of methanol, ethanol, and acetaldehyde**

Methanol ($CH_3OH$) is the simplest alcohol. Its IE is 10.84 eV, higher than the energy of the photons emitted from the krypton lamp. A weak signal of protonated methanol was observed when 4.6 ppmv methanol was sampled. Figure 2(A) shows the obtained LPPI mass spectrum of 4.6 ppmv methanol in nitrogen. The mass peaks at m/z 19, 37, and 55 correspond to $H_3O^+$, $(H_2O)_2H^+$, and $(H_2O)_3H^+$, respectively. The mass peaks at m/z 33, 51 and 65 are assigned to $(CH_3OH)H^+$, $(CH_3OH \cdot H_2O)H^+$ and $(CH_3OH)_2H^+$, respectively. The moderate mass peak at m/z 47 is assigned to ethanol, the impurity in the methanol reagent. The peak intensities of $(CH_3OH)H^+$ and $(CH_3OH)_2H^+$ are 559 and 171 counts, respectively. It is reported in the literature that dimers of methanol $(CH_3OH)_2$ with IE equal to 9.72 eV coexist with methanol monomers under ambient conditions and that protonated methanol is generated from the dissociation of $(CH_3OH)_2^+$.[28-29] Figure 2(B) shows the LPPI mass spectrum of 4.6 ppmv methanol doped with 291 ppmv $CH_2Cl_2$. The signal intensities of the mass peaks at m/z 33 and 65 reach $1.48 \times 10^5$ and $6.06 \times 10^4$ counts, respectively. The signal intensity of protonated methanol was averagely enhanced by 197.8 ± 18.8 times, measured from three independent measurements.

The IE of ethanol ($C_2H_5OH$) is 10.48 eV, meaning it can be photoionized by the photons emitted from the krypton lamp (10.6 eV, 20%). Figure 3(A) shows the LPPI mass spectrum of 1.6 ppmv ethanol in nitrogen. As well as ions resulting from water and water clusters, mass peaks at m/z 45, 47, and 93 are assigned to ions produced from ethanol, i.e. $C_2H_5O^+$ (551 counts), $(C_2H_5OH)H^+$ (1923 counts), and $(C_2H_5OH)_2H^+$ (222 counts). The mass peak of protonated ethanol was the strongest peak. After doping with 291 ppmv $CH_2Cl_2$, the intensities of the mass peaks at m/z 47 and 93 shown in Figure 3(B) increased to $1.61 \times 10^5$ and $2.21 \times 10^4$ counts, respectively. The signal intensity of protonated ethanol was averagely enhanced by 87.3 ± 7.8 times, measured from three independent measurements. The mass peak at m/z 45 slightly increased to 2765 counts, while the mass peaks at m/z 29 ($1.54 \times 10^4$ counts) and 65 ($1.80 \times 10^4$ counts) assigned to $C_2H_5^+$ and

$(C_2H_5OH \cdot H_2O)H^+$ appeared.

Acetaldehyde ($C_2H_4O$) is one of the most important aldehydes; it occurs widely in nature and is produced industrially on a large scale. The IE of acetaldehyde is 10.23 eV. Figure 4(A) shows the LPPI mass spectrum of 0.66 ppmv acetaldehyde in pure nitrogen. The mass peaks at m/z 45 and 61 are assigned to $(C_2H_4O)H^+$ (1290 counts), and $(C_2H_3O \cdot H_2O)^+$ (1307 counts), respectively. The molecular ion of acetaldehyde was not observed. Protonated acetaldehyde was the dominant ion. Figure 4(B) shows the LPPI mass spectrum of 0.66 ppmv acetaldehyde in nitrogen doped with 291 ppmv $CH_2Cl_2$. The signal intensity of protonated acetaldehyde (m/z 45) increased to $7.04 \times 10^4$ counts, while the signal at m/z 61 slightly increased to 2107 counts. The signal intensity of protonated acetaldehyde was averagely enhanced by $93.5 \pm 35.5$ times, measured from three independent measurements. Additionally, a mass peak at m/z 63 assigned to $(C_2H_4O \cdot H_2O)H^+$ ($1.71 \times 10^4$ counts) appeared.

Benzene ($C_6H_6$) is an important chemical and atmospheric pollutant. Its IE is 9.24 eV, lower than the energy of VUV photons emitted from the krypton lamp. Benzene and its derivatives have excellent photoionization efficiencies under illumination of a krypton VUV lamp. Figure 5(A) shows the LPPI mass spectrum of 0.42 ppmv benzene. The mass peak at m/z 78 is assigned to $^{12}C_6H_6^+$ ($6.42 \times 10^4$). Figure 5(B) shows the LPPI mass spectrum of 0.42 ppmv benzene in nitrogen doped with 291 ppmv $CH_2Cl_2$. The intensities of the mass peak at m/z 78 decreased by ~14% to $5.54 \times 10^4$ counts. The fluctuation of the signal intensities at m/z 78 was observed in separate experiments. No signal enhancement at m/z 79 (protonated benzene) was observed in all experiments.

**Discussion**

Pure $CH_2Cl_2$ in LPPI almost does not generate noticeable ions as shown in Figure 1, which implies that the observed protonation enhancement is not attributed to charge exchange. In order to reveal the mechanism of protonation enhancement, the doping effects of $H_2$, $CH_4$, $CHCl_3$, and $CCl_4$ on the signals of methanol, ethanol, and acetaldehyde were also investigated. Among the four dopants, only $CHCl_3$ yielded a weaker enhancement on protonation of methanol, ethanol, and acetaldehyde compared with $CH_2Cl_2$. Under illumination of the krypton lamp, $CH_4$, $CHCl_3$, $CH_2Cl_2$, and $CCl_4$ have relatively strong absorption ($\sim 10^{-17}$ cm$^2$) and are excited to Rydberg states[30-31], while $H_2$ does not have absorption[32]. Shaw *et al.* reported that ion-pair states were observed in halogenated methanes excited by VUV light and ion pair states even existed below ionization potentials[33]. We speculate that $CHCl_3$ and $CH_2Cl_2$, excited by the krypton lamp, may form the ion-pair states of $[H^+-CCl_3^-]$ and $[H^+-CHCl_2^-]$, which facilitate protonation. Other dopants including $H_2$, $CH_4$, and $CCl_4$ do not meet the combined conditions of formation of ion-pair states and existence of H atoms.

Table 1 lists IEs, PAs, molecular dipole moments, H-bond formation possibilities, and protonation enhancements of the compounds investigated. It is very interesting that protonation of benzene and self-protonation of dichloromethane were not observed in the experiment, while water and other three organics had significant protonation enhancements. The difference observed in protonation enhancements cannot be addressed simply by proton affinities or molecular dipole moments of the compounds. It is enlightening that the observed protonation enhancements of the compounds are coincident with their abilities to form a H bond as a H acceptor as shown in Table 1. The four compounds, water, methanol, ethanol, and acetaldehyde, are all capable of forming a H bond as a H acceptor, while benzene and dichloromethane are not. These phenomena may imply that the compounds are not protonated by free protons or protonated molecules. Based on experimental observations and the analyses above, we speculate that the following process might take place during $CH_2Cl_2$ doping:

$$CH_2Cl_2 + h\nu \rightarrow [H^+-CHCl_2^-] \quad (5)$$

$$A + [H^+-CHCl_2^-] \rightarrow [A-H^+-CHCl_2^-] \quad (6)$$

$$[A-H^+-CHCl_2^-] \rightarrow AH^+ + CHCl_2^- \quad (7)$$

where $[H^+-CHCl_2^-]$ represents an ion-pair state, $[A-H^+-CHCl_2^-]$ sketches a complex formed via a H bond, and A denotes analyte molecules, i.e. molecules of water, methanol, ethanol, and acetaldehyde. The proposed scenario of protonation enhancement is as follows: 1. $CH_2Cl_2$ excited by VUV light transforms into an ion-pair state ($[H^+-CHCl_2^-]$); 2. The analyte molecule collides with $[H^+-CHCl_2^-]$ and forms a complex $[A-H^+-CHCl_2^-]$ via a H bond; 3. The detachment of the proton from $CH_2Cl_2$ leads to the formation of a protonated analyte molecule ($AH^+$) and $CHCl_2^-$. This hypothesis rationalizes all the experimental observations. To the best of our knowledge, protonation via collision with excited-state molecules has not yet been documented. The heat of reaction ($\Delta_r H°$) of deprotonation of $CH_2Cl_2$ ($CH_2Cl_2 = CHCl_2^- + H^+$) is ~16.3 eV.[34] Considering the photon energy of VUV light (~10 eV) and PAs of analyte molecules (in the range of 7–9 eV),[35] the total process of Reactions 5 to 7 is exothermic for most VOCs. Though the authenticity and intrinsic mechanism of the process still needs further elaborate investigation, the observed phenomenon opens a new prospect for the improvement of the detection efficiency of VUV photoionization.

**Methods**

The experimental setup has been described in detail elsewhere.[25] Briefly, it mainly consisted of a 120 L test chamber and a LPPI mass spectrometer.

The 120 L test chamber was mainly built with an open-head stainless steel drum and covered with a thin Tedlar bag to ensure one atmospheric pressure during sampling. A stainless steel fan driven by a magnetic field was placed at the bottom of the test chamber to ensure quick mixing. Nitrogen was used as the buffer gas. An oil-free pump was used as the drain pump. Two mass flow controllers were used for gas samples. All experiments were performed under ambient atmospheric pressure and room temperature.

The LPPI mass spectrometer was recently developed in our laboratory. It characterizes with a LPPI source with an optical baffle and short reflectron time-of-flight mass spectrometer. The body of the LPPI source was a cylindrical stainless steel cavity 6 mm in diameter and 35 mm in length. A radio frequency-driven krypton VUV lamp was used as the VUV light source and coupled to the cylindrical stainless steel cavity with an $MgF_2$ window. The optical baffle was placed at the exit of the photoionization source to prevent the VUV light entering the mass spectrometer. The LPPI source was passivated with ~600 ppm $CH_2Br_2$ under illumination of VUV light for ~8 hours to suppress photoelectron formation in the experiment. The krypton lamp was laboratory-assembled and emitted VUV photons with energies of 10.0 eV (~80%) and 10.6 eV (~20%). The sample gas was introduced into the photoionization source and controlled by a needle valve. The sample flow was maintained at ~1 $cm^3$ $s^{-1}$. The pressure in the photoionization source was 500–1000 Pa. The mass spectrometer was a simple V-shaped time-of-flight mass spectrometer with a free-field flight distance of 460 mm. The cycle time of detection was 10 s.

In the experiments, a small amount of bottle-contained chemical was first injected into the test chamber. Then, 100 µL $CH_2Cl_2$ was injected into the test chamber and the mass spectra were subsequently acquired after each injection. The amount of methanol, ethanol, acetaldehyde, and benzene injected into the nitrogen-filled test chamber was 1.0, 0.5, 0.5, and 0.2 μL, respectively. The resulting mixing ratios for the prepared gases were 4.6, 1.6, 0.66, and 0.42 ppmv, respectively. The injection of 100 μL $CH_2Cl_2$ resulted in 291 ppmv in the mixing ratio.

In this study, methanol (A. R., Sinopharm), ethanol (A. R., Sinopharm), acetaldehyde (40% in water, Sinopharm), benzene (A. R., Beijing Shiji), $CH_2Cl_2$ (HPLC grade, Cleman Chemical), $CHCl_3$ (A. R., Beijing Shiji), and $CCl_4$ (A. R., Sinopharm) were used. High-purity nitrogen (>99.99%), hydrogen (>99.999%), and methane (>99.9%) were purchased from Beijing Huayuan Gas Co., Ltd.

**Acknowledgements**

This study was supported by the Strategic Priority Research Program of the Chinese Academy of Sciences (XDB05040501).

**Author Contributions**

J. S. conceived the idea, conducted the experiment, and wrote the paper. Y. Z., C. X., and Z. L. conducted the experiment; W. S. did the preliminary study which indicated the possibility of protonation enhancement. B. Y. and X. Z. were involved in preparing the manuscript and supplying chemicals. P. Z. and P. M. managed the experimental setup.

**Additional Information**

**Competing financial interests:** The authors declare no competing financial interests.

| Compounds | Ionization Energy [a] (eV) | Proton Affinity [a] (kJ/mol/eV) | Molecular Dipole Moment[b] ($10^{-30}$ C·m) | H-bond Formation Possibilities as H Acceptor | Protonation Enhancement by $CH_2Cl_2$ [c] (times) |
|---|---|---|---|---|---|
| $H_2O$ | 12.62 | 691/7.22 | 6.2 | Yes | 481.7±122.4 |
| $CH_3OH$ | 10.84 | 754.3/7.82 | 5.5 | Yes | 197.8±18.8 |
| $C_2H_5OH$ | 10.48 | 776.4/8.05 | 5.7 | Yes | 87.3±7.8 |
| $C_2H_4O$ | 10.23 | 768.5/7.97 | 8.3 | Yes | 93.5±35.5 |
| $C_6H_6$ | 9.24 | 750.4/7.78 | 0 | No | 0 |
| $CH_2Cl_2$ | 11.33 | 628/6.51 | 6.0 | No | 0 |

Table 1. Ionization energies (IEs), proton affinities (PAs), molecular dipole moments, and H-bond formation possibilities, and protonation enhancements of the compounds investigated. [a]http://webbook.nist.gov/. [b]http://www.kayelaby.npl.co.uk/. [c]obtained from three independent measurements.

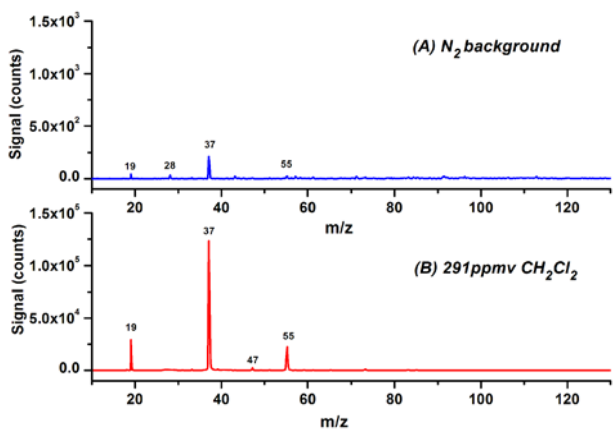

Figure 1. LPPI mass spectra of $N_2$ before (A) and after (B) doping with 291 ppmv $CH_2Cl_2$.

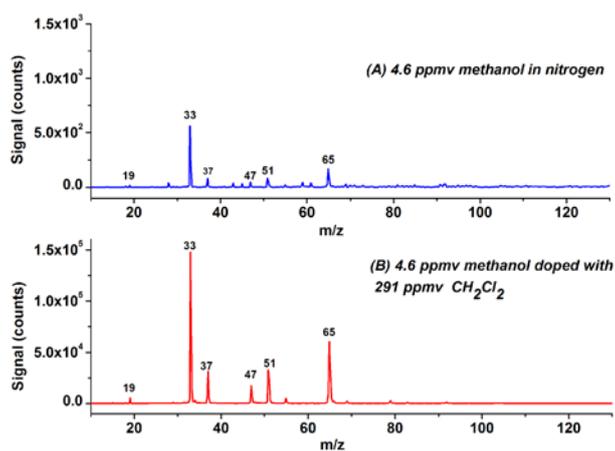

Figure 2. LPPI mass spectra of 4.6 ppmv methanol before (A) and after (B) doping with 291 ppmv $CH_2Cl_2$.

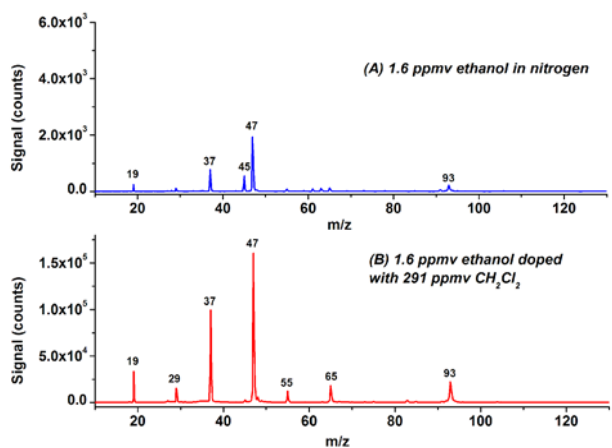

Figure 3. LPPI mass spectra of 1.6 ppmv ethanol before (A) and after (B) doping with 291 ppmv $CH_2Cl_2$.

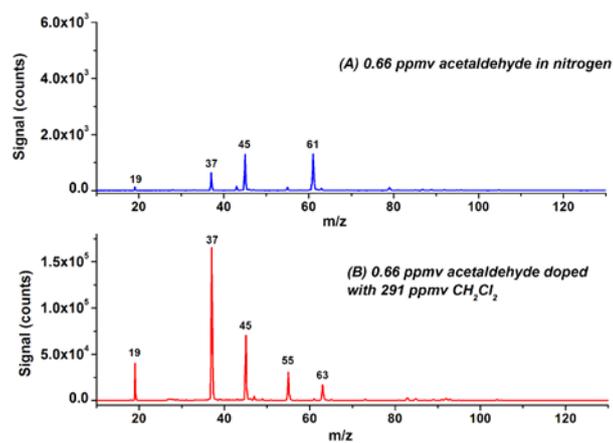

Figure 4. LPPI mass spectra of 0.66 ppmv acetaldehyde before (A) and after (B) doping with 291 ppmv $CH_2Cl_2$.

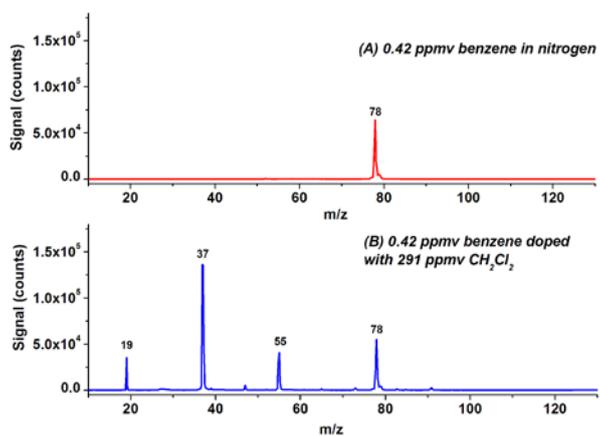

Figure 5. LPPI mass spectra of 0.42 ppmv benzene before (A) and after (B) doping with 291 ppmv $CH_2Cl_2$.